\documentstyle{ioplppt}
\begin{document}
\jl{1}
\title{Retrieval Phase Diagrams of  Non-monotonic Hopfield Networks 
}[Non-monotonic Hopfield Networks] 
\author{Jun-ichi Inoue} 
\address{
Department of Physics, Tokyo Institute of Technology,
Oh-okayama, Meguro-ku, Tokyo 152, Japan}
\begin{abstract}We investigate the retrieval phase diagrams of an asynchronous 
 fully-connected attractor network with non-monotonic transfer function by 
means of a mean-field approximation. 
We find for the noiseless zero-temperature case that this non-monotonic Hopfield network can store 
more patterns than a network with monotonic transfer function investigated by Amit {\it et al}. Properties of retrieval phase diagrams 
of non-monotonic networks agree with the results obtained by 
Nishimori and Opris  who treated synchronous networks. 
We  also  investigate the optimal storage capacity 
of the non-monotonic Hopfield model with state-dependent synaptic couplings 
introduced by Zertuche {\it et al}. 
We show that the non-monotonic Hopfield model with state-dependent 
synapses stores more patterns 
than the  conventional Hopfield model. 
Our formulation can be easily extended  to a general transfer function.
\end{abstract}
\maketitle
\section{Introduction}
Statistical mechanical approaches were successful  
for investigation of equilibrium 
 properties of associative memories or attractor networks.
The Hopfield model \cite{Hop1}\cite{Hop2} which updates  
its state asynchronously was investigated 
from statistical mechanical point of view 
by Amit {\it et al} \cite{Amit}\cite{Amit2}, and a lot of interesting 
features were found.
One of the main issues about the Hopfield model 
as an associative memory device is the critical storage capacity.
Amit {\it et al} \cite{Amit} showed that 
the Hebbian learning in the Hopfield model 
leads to the  optimal storage capacity ${\alpha}_{c}=p/N=0.138$, where 
$p$ is the number of embedded patterns and $N$ is the number of neurons. 
Fontanari and K$\ddot{\rm o}$berle \cite{Fon} 
extended the method of Amit {\it et al} 
to the synchronous  networks and showed that the capacity remains the same 
${\alpha}_{c}=0.138$ 
and derived finite-temperature properties of synchronous networks.
On the other hand, we can not obtain 
information on the dynamical process of retrieval  
by equilibrium statistical mechanics. 

Amari and Maginu \cite{Amari} proposed a signal-to-noise ratio analysis 
to investigate the dynamical properties of synchronous networks.
They divided time-dependent local field 
$h_{i}^{t}=(1/N)\sum_{j}\sum_{\mu}{\xi}_{i}^{\mu}{\xi}_{j}^{\mu}
{\sigma}_{j}(t)$ into 
a signal part $m_{t}$,  which corresponds to 
the first (${\mu}=1$) term in the summation over 
${\mu}$,  and a noise contribution 
$N_{i}^{t}$ corresponding to the rest (${\mu}{\neq}1$). 
They assumed  that the time-dependent noise term obeys 
the Gaussian distribution during the dynamical processes  and showed 
that the capacity is ${\alpha}_{c}=0.159$.
Nishimori and Ozeki \cite{Nishi2} pointed out by Monte-Carlo simulations that 
the assumption of a Gaussian distribution of the noise term is valid at 
least within statistical uncertainties if the final retrieval 
is successful. 
And they extended the Amari-Maginu theory 
to the network which updates its state stochastically and investigated 
the properties of the Hopfield network at finite temperatures. 
The phase diagram obtained as the equilibrium limit of 
the extended Amari-Maginu dynamics is very similar to the phase diagram 
of Amit {\it et al} \cite{Amit}\cite{Amit2}.

The limitation of storing patterns in Hopfield networks comes  
 mainly from the 
Hebbian interactions $J_{ij}=(1/N)\sum_{\mu=1}^{p}{\xi}_{i}^{\mu}{\xi}_{j}^{\mu}
$.
In fact, Gardner \cite{Gar} \cite{Gar2} \cite{Gar3} showed, 
in her pioneering papers, that 
the optimal storage capacity ${\alpha}_{c}$ is $2$ for the general 
interaction $J_{ij}$.
Many attempts have been made to increase 
the storage capacity of the Hopfield model to Gardner's limit ${\alpha}_{c}=2$
 by taking more complex synapses.
Recently Zertuche {\it et al} \cite{Ze} studied the storage capacity of 
the Hopfield model with state-dependent 
synapses by introducing a 
threshold parameter ${\eta}$.  
This parameter ${\eta}$ determines which patterns contribute to the 
synapses. This synapse can be written 
as $J_{ij}=(1/N)\sum_{\mu}
{\xi}_{i}^{\mu}{\xi}_{j}^{\mu}{\Theta}({m_{\mu}}^{2}-{\eta}^{2}/N)$. 
In their model, only patterns 
whose correlation with the state of the networks is
 greater or equal to the 
threshold are left finite to give a Hebbian contribution to the 
synapses. 
The capacity of the Hopfield network with this type of synapses is found to  
increase  ${\alpha}_{c}$ from $0.138$ to $0.171$ at $T=0$ and ${\eta}=1.0$.

Nishimori and Opris \cite{Nishi1} investigated the retrieval 
properties of an associative memory with a general 
transfer function using the extended Amari-Maginu theory \cite{Nishi2}. 
They obtained the optimal storge capacity 
for the non-monotonic transfer function by taking the  equilibrium 
limit of the recursion relation of the Amari-Maginu dynamics and 
showed that networks with non-monotonic transfer functions
 yield an enhanced 
memory capacity than the conventional monotonic relation. 
This property of non-monotonic neural networks was also pointed out by 
Morita {\it et al} \cite{Mori} by Monte-Carlo simulation before 
Nishimori and Opris \cite{Nishi1}.
The reason why the optimal storage capacity of non-monotonic 
transfer function increases is that a weak value of the total input to a neuron
implies a confused state and an inverted output  
for a weak input might works as a trial toward an improved retrieval. 

In this paper we investigate the retrieval phase diagram of Hopfield networks 
which update asynchronously and have  a non-monotonic transfer function by 
a mean-field theory of statistical mechanics proposed by Geszti \cite{Ge}\cite{Her}. In section 2 we show the formulation of the mean-field approximation 
to the asynchronous Hopfield networks with a non-monotonic transfer function 
and equation of state are derived. In section 3 we extend our formulation to the networks with 
a general type of transfer function. 
In section 4 we study the performance of non-monotonic Hopfield networks 
when their synapses depend  
on the state of networks using 
the method  proposed by Zertuche {\it et al} \cite{Ze}. 
In section 5 we compare the results of our calculations 
with  the results obtained by Nishimori and Opris \cite{Nishi1}.
 
\section{Equations of state}
Most of the investigations which discussed equilibrium properties 
of fully-connected Hopfield networks by statistical mechanics 
were restricted to networks with equilibrium free energy.  
In order to show the existence of such a free energy, we must 
make it sure 
that the synaptic couplings are symmetric and 
the transfer function is monotonic. 
Our model in this paper has symmetric 
couplings, but the transfer function is non-monotonic. 
In order to overcome this difficulty we use the mean-field approximation 
as follows.

Let us suppose that the $i$th neuron updates its state according to the next 
 probability \cite{Nishi2}
\begin{eqnarray}
{\rm Prob}({\sigma}_{i}(t+1))=\frac{1}{2}\left[1+{\sigma}_{i}
(t+1)f(h_{i}^{t})\right]
\end{eqnarray}
where ${\sigma}_{i}={\pm}1$, ${\xi}_{i}^{\mu}={\pm}1$, and 
the local field to the $i$th neuron $h_{i}^{t}$ is defined as 
\begin{eqnarray}
h_{i}^{t}=\frac{1}{N}\sum_{j{\neq}i}\sum_{\mu}{\xi}_{i}^{\mu}{\xi}_{j}^{\mu}
{\sigma}_{j}(t)
\end{eqnarray}
Then we can calculate the average value of ${\sigma}_{i}(t+1)$ as
\begin{eqnarray}
<{\sigma}_{i}(t+1)>=(+1){\times}\frac{1}{2}\left[1+f(h_{i}^{t})\right]
+(-1){\times}\frac{1}{2}\left[1-f(h_{i}^{t})\right]\nonumber \\
\mbox{}=f(h_{i}^{t})\hspace{2.6in}
\end{eqnarray}
For the equilibrium state we obtain the equation of state 
by mean field approximation
 \cite{Ge} as 
\begin{eqnarray}
<{\sigma}_{i}>=f\left(\frac{1}{N}\sum_{j}\sum_{\mu}{\xi}_{i}^{\mu}{\xi}_{j}^{\mu}
<{\sigma}_{j}>\right)
\end{eqnarray}
In this section  we choose the function $f(x)$ as 
\begin{eqnarray}
f(x)={\tanh}\left(-{\beta}(x+a)\right)+{\tanh}\left(-{\beta}(x-a)\right)
+{\tanh}\left({\beta}x\right)
\end{eqnarray}
where $a$ is a positive constant and ${\beta}$ is a parameter related to 
the synaptic noise. 
This non-monotonic transfer function reduces to the form 
in Figure 1 in the limit 
${\beta}=1/T{\rightarrow}\infty$.
Then equation (4) can be rewritten explicitly as 
\begin{eqnarray}
<{\sigma}_{i}>={\tanh}\left[-\frac{\beta}{N}\sum_{j}\sum_{\mu}
{\xi}_{i}^{\mu}{\xi}_{j}^{\mu}<{\sigma}_{j}>-{\beta}a\right] \nonumber \\
\mbox{}+{\tanh}\left[-\frac{\beta}{N}\sum_{j}\sum_{\mu}
{\xi}_{i}^{\mu}{\xi}_{j}^{\mu}<{\sigma}_{j}>+{\beta}a\right] \nonumber \\
\mbox{}+{\tanh}\left[\frac{\beta}{N}\sum_{j}\sum_{\mu}
{\xi}_{i}^{\mu}{\xi}_{j}^{\mu}<{\sigma}_{j}>\right]
\end{eqnarray}
We introduce the overlap between the equilibrium 
state of the network $<{\sigma}_{i}>$ and an 
 embedded pattern ${\nu}$ as follows
\begin{eqnarray}
m_{\nu}{\equiv}\frac{1}{N}\sum_{i}{\xi}_{i}^{\nu}<{\sigma}_{i}>
\end{eqnarray}
Using this overlap parameter, we may rewrite (6) as
\begin{eqnarray}
m_{\nu}=\frac{1}{N}\sum_{i}{\xi}_{i}^{\nu} \hspace{4.2in}\nonumber \\
\mbox{}{\times}\left\{{\tanh}\left[{\beta}
(-\sum_{\mu}{\xi}_{i}^{\mu}m_{\mu}+a)\right]+{\tanh}
\left[{\beta}(-\sum_{\mu}{\xi}_{i}^{\mu}m_{\mu}-a)\right]
+{\tanh}\left[{\beta}\sum_{\mu}{\xi}_{i}^{\mu}\right ]\right\}
\end{eqnarray}
We divide the term $\sum_{\mu}{\xi}_{i}^{\mu}m_{\mu}$ 
appearing above into three parts: the first for ${\mu}=1({\neq}{\nu})$ 
which corresponds to the retrieved state, the second corresponding to 
the term ${\mu}={\nu}$, and the rest.
 Then we get
\begin{eqnarray}
m_{\nu}=-\frac{1}{N}\sum_{i}{\xi}_{i}^{\nu}{\xi}_{i}^{1}
{\tanh}\left[{\beta}(m_{1}+{\xi}_{i}^{\nu}{\xi}_{i}^{1}
m_{\nu}+\sum_{\mu{\neq}1,\nu}{\xi}_{i}^{\mu}{\xi}_{i}^{1}m_{\mu}-a)\right] \nonumber \\
\mbox{}-\frac{1}{N}\sum_{i}{\xi}_{i}^{\nu}{\xi}_{i}^{1}
{\tanh}\left[{\beta}(m_{1}+{\xi}_{i}^{\nu}{\xi}_{i}^{1}
m_{\nu}+\sum_{\mu{\neq}1,\nu}{\xi}_{i}^{\mu}{\xi}_{i}^{1}m_{\mu}+a)\right] \nonumber \\
\mbox{}+\frac{1}{N}\sum_{i}{\xi}_{i}^{\nu}{\xi}_{i}^{1}
{\tanh}\left[{\beta}(m_{1}+{\xi}_{i}^{\nu}{\xi}_{i}^{1}
m_{\nu}+\sum_{\mu{\neq}1,\nu}{\xi}_{i}^{\mu}{\xi}_{i}^{1}m_{\mu})\right]
\end{eqnarray}
Here, $m_{1}$ is of order $1$, and the summation $\sum_{\mu{\neq}1,\nu}
{\xi}_{i}^{\mu}{\xi}_{i}^{1}m_{\mu}$ is also of order $1$, while 
the term ${\xi}_{i}^{\nu}{\xi}_{i}^{1}m_{\nu}$ is 
much smaller, ${\cal O}(1/\sqrt{N})$.
Then we may 
 expand the function ${\tanh}$ appearing in (9) to 
 first order of ${\xi}_{i}^{\nu}
{\xi}_{i}^{1}m_{\nu}$. 
The terms  $\sum_{\mu{\neq}1,{\nu}}{\xi}_{i}^{\mu}{\xi}_{i}^{1}
m_{\mu}$ appearing in ${\tanh}$ may be regarded as Gaussian variables with 
mean zero and variance $\sum_{\mu{\neq}1,{\nu}}
{m_{\mu}}^{2}{\equiv}{\alpha}r$ \cite{Ge}.
Under this approximation, we may replace the summation $(1/N)\sum_{i}$ by a
Gaussian integral. We next square the 
equation (9) to calculate $r=\sum_{{\nu}{\neq}1}{m_{\nu}}^{2}/{\alpha}$.
Following the procedure introduced by Gesti \cite{Ge}, we obtain the 
equations of state in the limit of $N{\rightarrow}\infty$ as follows
\begin{eqnarray}
r=\left[1-{\beta}(q_{+}+q_{-}-q-1)\right]^{-2}
\end{eqnarray}
and 
\begin{eqnarray}
m=-\int{Dz}{\tanh}[{\beta}(m+\sqrt{{\alpha}r}z-a)] 
\nonumber \\ 
\mbox{}-\int{Dz}{\tanh}[{\beta}(m+\sqrt{{\alpha}r}z+a)] 
\nonumber \\ 
\mbox{}+\int{Dz}{\tanh}[{\beta}(m+\sqrt{{\alpha}r}z)] 
\end{eqnarray}
where we set $m_{1}=m$ and introduced the Edward-Anderson \cite{Ed} 
like order parameters 
\begin{eqnarray}
q\,{\equiv}\,\int{Dz}{\tanh}^{2}\left[{\beta}(m_{1}+\sqrt{{\alpha}r}z)\right] \hspace{.24in}
\end{eqnarray}
\begin{eqnarray}
q_{\pm}\,{\equiv}\,\int{Dz}{\tanh}^{2}\left[{\beta}(m+\sqrt{{\alpha}r}z\,{\pm}\,a)\right]
\end{eqnarray}
In the zero temperature limit, these order parameters can be rewritten as 
\begin{eqnarray}
q=1-\sqrt{\frac{2}{\pi\alpha{r}{\beta}^{2}}}\,{\exp}
\left(-\frac{{m}^{2}}{2{\alpha}r}\right)\hspace{.3in}
\end{eqnarray}
\begin{eqnarray}
q_{\pm}=1-\sqrt{\frac{2}{\pi\alpha{r}{\beta}^{2}}}\,{\exp}
\left(-\frac{(m{\pm}a)^{2}}{2{\alpha}r}\right)
\end{eqnarray}
Then equations of state lead to 
\begin{eqnarray}
r=\left[1+\sqrt{\frac{2}{\pi{\alpha}r}}\left\{
\exp\left(-\frac{(m+a)^{2}}{2{\alpha}r}\right)
+\exp\left(-\frac{(m-a)^{2}}{2{\alpha}r}\right)
+\exp\left(-\frac{m^{2}}{2{\alpha}r}\right)\right\}\right]^{-2}
\end{eqnarray}
and
\begin{eqnarray}
m=-{\rm erf}\left(\frac{m-a}{\sqrt{2{\alpha}r}}\right)
-{\rm erf}\left(\frac{m+a}{\sqrt{2{\alpha}r}}\right)
+{\rm erf}\left(\frac{m}{\sqrt{2{\alpha}r}}\right)
\end{eqnarray}
where
\begin{eqnarray}
{\rm erfc}(x){\equiv}1-{\rm erf}(x)=\frac{2}{\sqrt{\pi}}
\int_{x}^{\infty}dt\,{\exp}(-t^{2})
\end{eqnarray}
For $a{\rightarrow}\infty$ one recovers the equations of state 
of the Hopfield model at $T=0$ obtained by Amit et al 
\cite{Amit}\cite{Amit2}.

The neural network is useful  as an associative memory 
as long as the mean field equations (10) and (11) have a solution of the form 
$\mbox{\boldmath $m$}_{\mu}=(m,0,{\cdots},0,0)$ with $m{\neq}0$.  
At $T=0$, there exist metastable states which are 
highly correlated with particular embedded 
patterns as long as $p=
{\alpha}N<{\alpha}_{c}N$. 
We solved the equation (16) and (17) numerically and 
obtained  the critical capacity ${\alpha}_{c}$. This result is plotted 
as a function of the parameter $a$ in figure 2. 
In this phase diagram, the region R denotes the retrieval phase and 
the region N/R means the non-retrieval 
phase where the self-consistent 
equations (16) and (17) 
do not have a non-zero solution of $m$. 
From this result, we see that Hopfield networks 
with non-monotonic transfer function store more 
patterns than the networks with monotonic transfer function \cite{Amit}:
 ${\alpha}_{c}$  has the maximum value $0.211$ at 
$a=1.77$.
The shape of the ${\alpha}_{c}(a)$ curve has similar properties, 
in the following sense, 
with that of Nishimori and Opris \cite{Nishi1} 
who calculated this critical curve by the equilibrium 
relation of Amari-Maginu dynamics for synchronous
networks:
\begin{itemize}
\item There exists a certain value of $a$ that maximizes ${\alpha}_{c}$.
\item ${\alpha}_{c}$ approaches the monotonic value (0.138 
in the present case) in the limit $a{\rightarrow}\infty$.
\end{itemize}
An interesting observation is that an iterative solution of 
the self-consistent equations (16) and (17) showed  
oscillatory behavior in a restricted region 
around $a=0$ and ${\alpha}=0$ in the 
phase diagram. 
In consideration of similar observations from dynamical treatments 
\cite{Nishi1}, 
such a phase should be characterized by time development of 
the overlap 
$m_{t}$ in the sense that $m_{t+1}>0$ if $m_{t}<0$ and $m_{t+1}<0$ 
if $m_{t}>0$. 
However, it should be warned that 
as we treat the static properties 
of the non-monotonic 
Hopfield model, we can not extract the 
dynamical behavior of $m_{t}$ from our 
formulation, strictly speaking. 

Next, 
in order to get the ${\alpha}$-$T$ phase diagram, 
we solved  the finite-temperature self consistent equations (10) and 
(11). 
For the case of $a=1.80$ we plotted the ${\alpha}$-$T$ curve in figure 3.
We have found that the transition from the (normal) retrieval phase 
to the spin glass phase is of first order.

\section{Extension to a general transfer function}
In this section we show that our formulation can 
be extended to the networks with a general 
transfer function \cite{Nishi1}.
The mean field equation of state for a general transfer function 
$f$ is already given in (4). 
Equation (9) of the parameter $m_{\nu}$ 
can be rewritten for this general transfer function as follows.
\begin{eqnarray}
m_{\nu}=\frac{1}{N}\sum_{i}{\xi}_{i}^{\nu}f(\sum_{\mu}{\xi}_{i}^{\mu}m_{\mu})
\end{eqnarray}
Here we have included the effect of the control parameter $\beta$, 
corresponding to the thermal noise, in the general function $f$.
We should not forget that the absolute value of the function 
$f$ does not exceed $1$ 
because otherwise probabilistic interpretation 
(1) does not make sense.
For the general transfer function $f$, we can obtain the 
equations of state in the same way as in the 
previous section. 
The result is 
\begin{eqnarray}
r=\frac{Q_{0}}{[1-Q]^{2}}
\end{eqnarray}
and 
\begin{eqnarray}
m=\int\,{Dz}\,f(m+\sqrt{{\alpha}r}z)
\end{eqnarray}
where
\begin{eqnarray}
Q_{0}\,{\equiv}\,\int\,{Dz}\,f^{2}(m+\sqrt{{\alpha}r}z)
\end{eqnarray}
\begin{eqnarray}
Q\,{\equiv}\,\int\,{Dz}\,f'(m+\sqrt{{\alpha}r}z)
\end{eqnarray}
Note that by setting $f(x)={\tanh}({\beta}x)$ we 
recover the result by Amit {\it et al} \cite{Amit}.

We next show that our equations of state  
for the general transfer function are different 
from the result of 
Nishimori and Opris 
\cite{Nishi1}. They calculated the recursion relations of 
macro-variables $m_{t}$ and ${\sigma}_{t}$ 
(the latter being the measure of disturbance from non-retrieved patterns) 
by generalizing the Amari-Maginu type signal-to-noise 
ratio analysis \cite{Amari} to stochastic dynamical process 
 in the case of synchronous dynamics. 
Their result is
\begin{eqnarray}
m_{t+1}=\int{Dz}f(m_{t}+{\sigma}_{t}z)
\end{eqnarray}
\begin{eqnarray}
{\sigma}_{t+1}^{2}={\alpha}+2{\alpha}m_{t}m_{t+1}
h(m_{t},{\sigma}_{t})+{\sigma}_{t}^{2}h^{2}(m_{t},{\sigma}_{t})
\end{eqnarray}
where $h$ is defined by 
\begin{eqnarray}
h(m_{t},{\sigma}_{t})=\int{Dz}f'(m_{t}+{\sigma}_{t}z)
\end{eqnarray}
Taking the equilibrium limit $t{\rightarrow}\infty$ and 
setting ${\sigma}_{\infty}=\sqrt{{\alpha}r}$ and $m_{\infty}=m$, we get 
equations of state with respect to  $m$ and $r$ as 
\begin{eqnarray}
m=\int{Dz}f(m+\sqrt{{\alpha}r}z)
\end{eqnarray}
\begin{eqnarray}
r=\frac{1+{\alpha}m^{2}h(m,\sqrt{{\alpha}r})}
{\left[1-\left\{h(m,\sqrt{{\alpha}r})\right\}^{2}\right]}
\end{eqnarray}
where 
\begin{eqnarray}
h(m,\sqrt{{\alpha}r})=\int{Dz}f'(m+\sqrt{{\alpha}r}z)
\end{eqnarray} 
This is different from our result (20) and (21). 
We may suppose  that this difference comes from the difference between 
synchronous and  asynchronous dynamics. 
There is no {\it a priori} reason why the equilibrium 
properties of synchronous networks should 
coincide with those with asynchronous networks.

\section{State-dependent synapses}
As long as the number of 
embedded patterns satisfies $p{\ll}N$, 
the noise term $\sum_{\mu{\neq}1,{\nu}}
{\xi}_{i}^{\mu}{\xi}_{i}^{\mu}m_{\mu}$ appearing in the 
mean-field equation 
is of order ${\cal O}(1/N)$ 
and we can neglect this term. However, if $p={\alpha}N$ with 
${\alpha}$ finite, this same term becomes 
${\cal O}(1)$ and this contribution can 
not be neglected. 
These non-retrieved memories 
${\xi}^{\mu}$ (${\mu}{\neq}1,{\nu}$), which appear in 
 $\sum_{\mu{\neq}1,{\nu}}{\xi}_{i}^{\mu}{\xi}_{i}^{\mu}m_{\mu}$, 
prevent networks from retrieving the embedded pattern.
The storage capacity of the Hopfield 
model is limited by the contribution of a 
large number of weakly correlated patterns. 
For the conventional Hopfield model, patterns $\{{\xi}^{\mu}\}$ 
are stored by the Hebbian type synaptic interaction $J_{ij}=(1/N)
\sum_{\mu}{\xi}_{i}^{\mu}{\xi}_{j}^{\mu}$. 
Therefore, in order to exclude 
non-retrieved memories which have small overlaps 
with the state of the network, we should modify the synaptic interaction 
so that only patterns with large overlaps with state 
contribute to the Hebbian 
rule \cite{Ze}.
Our main interest in this section is to what degree 
the stability of the memorized states is improved 
by this state-dependent synaptic interaction \cite{Ze}\cite{Mat} and  
how many patterns are stored 
in the Hopfield networks with non-monotonic transfer function.

We use the next state-dependent synapses  by introducing a threshold 
${\eta}(\,{\geq}0\,)$ \cite{Ze}.
\begin{eqnarray}
J_{ij}=\frac{1}{N}\sum_{\mu}{\xi}_{i}^{\mu}{\xi}_{j}^{\mu}{\Theta}(
{m_{\mu}}^{2}-\frac{{\eta}^{2}}{N})
\end{eqnarray}
The factor of the step 
function ${\Theta}(x)$ means that an embedded  
pattern ${\xi}^{\mu}$ is excluded if the overlap 
between the pattern and network state $m_{\mu}$ is below a threshold  
${m_{\mu}}^{2}<{\eta}^{2}/N$. 
We expect that 
the performance of a network as an associative memory is improved by 
introducing this type of synapses with threshold ${\eta}>0$ 
to exclude the spurious memories disturbing retrieval.
We introduced the factor $1/N$  because $m_{\mu}$ 
is of order $1/\sqrt{N}$.
It is important to bear in mind that the conventional 
Hebb interaction $J_{ij}=(1/N)\sum_{\mu}{\xi}_{i}^{\mu}
{\xi}_{i}^{\mu}$ is recovered by setting ${\eta}=0$.
Using this coupling, 
we rewrite the mean field 
equation obtained in section 2 as follows.
\begin{eqnarray}
m_{\nu}=\frac{1}{N}\sum_{i}
{\xi}_{i}^{\nu}{\xi}_{i}^{1}{\tanh}[{\beta}(m+{\eta}_{i}^{\mu})] \hspace{2.0in}\nonumber \\
\mbox{}+{\beta}
m_{\nu}{\Theta}({m_{\nu}}^{2}-\frac{{\eta}^{2}}{N})\frac{1}{N}
\sum_{i}[1-{\tanh}^{2}{\beta}(m+{\eta}_{i}^{\mu})] \nonumber \\
\mbox{}-\frac{1}{N}\sum_{i}
{\xi}_{i}^{\nu}{\xi}_{i}^{1}{\tanh}[{\beta}(m+{\eta}_{i}^{\mu}-a)] \nonumber \\
\mbox{}-{\beta}
m_{\nu}{\Theta}({m_{\nu}}^{2}-\frac{{\eta}^{2}}{N})\frac{1}{N}
\sum_{i}[1-{\tanh}^{2}{\beta}(m+{\eta}_{i}^{\mu}-a)] \nonumber \\
\mbox{}-\frac{1}{N}\sum_{i}
{\xi}_{i}^{\nu}{\xi}_{i}^{1}{\tanh}[{\beta}(m+{\eta}_{i}^{\mu}+a)] \nonumber \\
\mbox{}-{\beta}
m_{\nu}{\Theta}({m_{\nu}}^{2}-\frac{{\eta}^{2}}{N})\frac{1}{N}
\sum_{i}[1-{\tanh}^{2}{\beta}(m+{\eta}_{i}^{\mu}+a)] 
\end{eqnarray}
where we introduced ${\zeta}_{i}^{\mu}$ as follows. 
\begin{eqnarray}
{\zeta}_{i}^{\mu}{\equiv}\sum_{\mu{\neq}1,\nu}{\xi}_{i}^{\mu}
{\xi}_{i}^{1}m_{\mu}{\Theta}({m_{\mu}}^{2}-\frac{{\eta}^{2}}{N})
\end{eqnarray}
This is the sum of a large number ($={\alpha}N$) of small 
terms (of order $1/N$). 
We now assume that the small contributions $m_{\mu}$ (${\mu}{\neq}1$) 
have identical Gaussian distributions centered at zero, with variance 
${\sigma}^{2}/N$.
Strictly speaking, this statement is not exact because $m_{\mu}$ 
are related through (31). 
Nevertheless we accept this approximation in this paper.
By the same arguments, 
${\zeta}_{j}^{\mu}$ is assumed to have 
a Gaussian distribution with variance ${\alpha}r$ and average zero, so that
\begin{eqnarray}
{\alpha}r={\ll}({\eta}_{i}^{\nu})^{2}{\gg}
\end{eqnarray}
We also introduce the Edwards-Anderson like order parameters \cite{Ed} as 
\begin{eqnarray}
q\,{\equiv}\,\frac{1}{N}\sum_{i}
{\tanh}^{2}{\beta}(m+{\eta}_{i}^{\mu})=\int{Dz}\,{\tanh}^{2}{\beta}(m+\sqrt{{\alpha}r}z)\hspace{0.6in}
\end{eqnarray}
\begin{eqnarray}
q_{\pm}\,{\equiv}\,\frac{1}{N}\sum_{i}
{\tanh}^{2}{\beta}(m+{\eta}_{i}^{\mu}{\pm}a)=\int{Dz}
\,{\tanh}^{2}{\beta}(m+\sqrt{{\alpha}r}z{\pm}a)
\end{eqnarray}
Using these parameters, (31) leads to 
\begin{eqnarray}
m_{\nu}\left[1-{\beta}(q_{+}+q_{-}-q-1)
{\Theta}({m_{\nu}}^{2}-\frac{{\eta}^{2}}{N})\right] \hspace{1.0in} \nonumber \\
\mbox{}=\frac{1}{N}\sum_{i}{\xi}_{i}^{\nu}{\xi}_{i}^{1}{\tanh}
[{\beta}(m+{\eta}_{i}^{\mu})] \hspace{0.5in}\nonumber \\
\mbox{}-\frac{1}{N}\sum_{i}{\xi}_{i}^{\nu}{\xi}_{i}^{1}{\tanh}
[{\beta}(m+{\eta}_{i}^{\mu}-a)] \hspace{0.5in}\nonumber \\
\mbox{}-\frac{1}{N}\sum_{i}{\xi}_{i}^{\nu}{\xi}_{i}^{1}{\tanh}
[{\beta}(m+{\eta}_{i}^{\mu}+a)] \hspace{0.5in} 
\end{eqnarray}
Squaring this expression and averaging it over 
the distribution of patterns, we get
\begin{eqnarray}
{\sigma}^{2}+\left\{[1-{\beta}(q_{+}+q_{-}-q-1)]^{2}\right\}r=1
\end{eqnarray}
Equations (32) and (33) lead to 
\begin{eqnarray}
{\alpha}r=p{\ll}{m_{\mu}}^{2}{\Theta}({m_{\mu}}^{2}
-\frac{{\eta}^{2}}{N}){\gg} 
=p\int_{-\infty}^{\infty}
\frac{dz}{\sqrt{2\pi{\sigma}^{2}/N}}
{\exp}(-\frac{Nz^{2}}{2{\sigma}^{2}})
z^{2}{\Theta}(z^{2}-\frac{{\eta}^{2}}{N})
\end{eqnarray}
Using the transformation $Nz^{2}/2{\sigma}^{2}=t$, we find 
\begin{eqnarray}
{\alpha}r=\frac{p}{N}(\frac{2}{\sqrt{\pi}}){\sigma}^{2}
\int_{{{\eta}^{2}}/{2{\sigma}^{2}}}
^{\infty}t^{-\frac{1}{2}}{\exp}(-t)dt 
={\alpha}\frac{2}{\sqrt{\pi}}{\sigma}^{2}{\Gamma}(\frac{3}{2},\frac{
{\eta}^{2}}{2{\sigma}^{2}})
\end{eqnarray}
The final expression of $r$ is 
\begin{eqnarray}
r=\frac{2}{\sqrt{\pi}}{\sigma}^{2}{\Gamma}(\frac{3}{2},\frac{
{\eta}^{2}}{2{\sigma}^{2}})
\end{eqnarray}
where ${\Gamma}$ is the incomplete gamma function defined as 
\begin{eqnarray}
{\Gamma}(z,p)=\int_{p}^{\infty}
{\exp}(-t)t^{z-1}dt
\end{eqnarray}
Another equation is obtained for $m$ by taking ${\mu}=1$ in (31) in the 
limit ${\beta}{\rightarrow}\infty$. This result agrees with 
(11) in section 2.
The term ${\beta}(q_{+}+q_{-}-q-1)$ appearing in (37) leads  
in the limit ${\beta}{\rightarrow}\infty$  to 
\begin{eqnarray}
C\,{\equiv}\,{\beta}(q_{+}+q_{-}-q-1) \hspace{3.1in}\nonumber \\
\mbox{}=-\sqrt{\frac{2}{\pi{\alpha}r}}\left\{
{\exp}(-\frac{(m+a)^{2}}{2{\alpha}r})+{\exp}
(-\frac{(m-a)^{2}}{2{\alpha}r})-{\exp}(-\frac{m^{2}}{2{\alpha}r})\right\}
\end{eqnarray}
Finally we have the equations of state as follows.
\begin{eqnarray}
{\sigma}^{2}+[(1-C)^{2}-1]r=1
\end{eqnarray}
\begin{eqnarray}
r=\frac{2}{\sqrt{\pi}}
{\sigma}^{2}{\Gamma}(\frac{3}{2},\frac{{\eta}^{2}}{2{\sigma}^{2}})
\end{eqnarray}
\begin{eqnarray}
m=-{\rm erf}(\frac{m-a}{\sqrt{2{\alpha}r}})-{\rm erf}
(\frac{m+a}{\sqrt{2{\alpha}r}})+{\rm erf}
(\frac{m}{\sqrt{2{\alpha}r}})
\end{eqnarray}
For simplicity we use a variable $k$ defined by 
\begin{eqnarray}
k=\frac{\sqrt{\pi}}{2}
\frac{1}{{\Gamma}(\frac{3}{2},\frac{{\eta}^{2}}{2{\sigma}^{2}})}
-1
\end{eqnarray}
Equations (43), (44) and (46) are written as  
\begin{eqnarray}
r=\frac{1}{(1-C)^{2}+k}
\end{eqnarray}
\begin{eqnarray}
{\sigma}^{2}=\frac{1+k}{(1-C)^{2}+k}
\end{eqnarray}
For ${\eta}=0$, one has $k=0$ and from (43), (44) and (46) one recovers the 
equations for the non-monotonic 
Hopfield model at $T=0$ discussed in the previous section.

We evaluated equations (45), (46) and (43) and obtained  
the optimal storage capacity 
${\alpha}_{c}$. We show the results for ${\alpha}_{c}$ 
as a function of $\eta$ for the case
$T=0, a=3.0$ in figure 4.  
 A similar result is plotted in figure 5 for the case of $T=0$ and $a=1.80$.
We also show the parameter-$a$ dependence 
of the capacity ${\alpha}_{c}$ for the cases  of ${\eta}=1.0,  0.8$ and  
$0.6$ in figure 6.
From this figure we see that as the threshold parameter ${\eta}$ 
increases, more patterns can be embedded by the modified Hebbian rule 
(30).
It is observed how the storage capacity of the non-monotonic 
Hopfield networks is improved as the value of ${\eta}$ increases 
 from ${\eta}=0$. Therefore 
using the neural networks with non-monotonic 
transfer function and state-dependent synapses, we can get an 
associative memory with the high-quality 
performance. 
We also show the overlap $m$ as a function of ${\alpha}$ 
for the case of 
$\eta=0.80$, $a=3.00$ and $T=0$, $\eta=0.80$, $a=1.60$ and $T=0$ 
 in figure 7 and figure 8 respectively. 
From these figures $m({\alpha})$ is seen to drop to zero 
discontinuously at the critical 
capacity ${\alpha}_{c}{\sim}0.155$ for $a=3.0$ and 
${\alpha}_{c}{\sim}0.257$ for $a=1.6$.

\section{Discussion}
We have investigated the retrieval phase diagrams  
by the mean field approximation \cite{Ge} in the Hopfield networks 
with asynchronous dynamics. 
Mean field approximation was extended to the general type of transfer function.
The result shows that a non-monotonic transfer function yields an 
enhanced memory capacity for $a$ around $0.211$. 
This confirms the claim of Morita {\it et al} \cite{Mori} who 
used the numerical simulation for synchronous dynamics 
and the result of Nishimori and 
Opris \cite{Nishi1} who used 
the equilibrium relation 
of the Amari and Maginu 
\cite{Amari} dynamics for synchronous
dynamics.
The properties of the phase 
diagram obtained in this paper qualitatively resemble 
those of the phase diagram of the synchronous neural networks.
It is interesting that our calculation for the 
asynchronous network also showed the enhancement of the capacity as in the 
synchronous case: The shape of the retrieval phase diagram in this paper 
is similar to that of Nishimori and Opris \cite{Nishi1}. 
A difference is that  
within our formulation 
of the asynchronous 
networks, the oscillatory phase (limit-cycle  
phase) found by Nishimori 
and Opris was not obtained clearly. This phase 
is characterized by the behavior of the 
dynamical order parameter $m_{t}$ which is 
$m_{t+1}>0$ if $m_{t}<0$ and $m_{t+1} < 0$ if $m_{t}>0$ 
in the range of $0<a<1$. 
As we used the equilibrium statistical mechanics to get the phase diagram 
of the non-monotonic Hopfield model, we can not draw definite 
conclusions about the 
dynamical order parameter $m_{t}$. 

For the non-monotonic Hopfield model, the property  of 
asynchronous dynamics is an open problem.  
  However, the oscillatory behavior during the 
process of recursion-type solution of equilibrium 
equations of state may be related 
to the dynamical oscillatory phase 
found in the same region of the phase diagram.

We extended  our formulation to the general transfer 
function in this paper. 
It is interesting to investigate whether 
the storage capacity is enhanced by a transfer function 
which has a different shape from the stepwise-type one.  
For the moment, we could not find better transfer functions than 
the stepwise-type one.
And we also showed that the Hopfield network 
with non-monotonic transfer function and state-dependent 
couplings can store large number of patterns. 
From these results one can confirm that 
the limit of an associative memory with the Hebb type interactions  
consists in the effects of spurious states each of which has a small 
correlation with embedded patterns.
 
We may be able to find a new type of network which shows better 
performance than the conventional one by introducing 
non-monotonic transfer function and state-dependent synapses.

The author thanks Professor Hidetoshi Nishimori for many useful discussions 
during this work.
He also thanks Dr. Tomoko Ozeki for many suggestions 
about dynamical properties of Hopfield networks and the Amari-Maginu theory.

\newpage
{\Large \bf Figure captions}\\
\\
{\bf Figure 1}\\
The stepwise-type non-monotonic transfer function.\\
{\bf Figure 2}\\
Phase diagram for the non-monotonic transfer function 
in figure 1 obtained by the mean field approximation. 
Retrieval is successful in the region ${\bf R}$ 
(normal retrieval) and unsuccessful in ${\bf N/R}$.
${\alpha}_{c}(a=\infty)=0.138$
(consistent with  Amit et al) and ${\alpha}_{c}(a=1.77)=0.211$
(maximum value).
\\
{\bf Figure 3}\\
The ${\alpha}$-$T$ phase diagram of the Hopfield model 
with non-monotonic transfer function ($a=1.80$).
Retrieval is successful in ${\bf R}$ (retrieval phase)
 and unsuccessful in ${\bf SG}$ (spin glass phase).
\\
{\bf Figure 4}\\
A slice of the  phase space for the non-monotonic 
networks with threshold ${\eta}$ at $T=0$ and $a=3.0$. 
The Hopfield value ${\alpha}_{c}=0.138$ 
is found at ${\eta}=0$; for a threshold ${\eta}$ equal to $1$ the 
optimal capacity increases to ${\alpha}_{c}=0.171$.
\\
{\bf Figure 5}\\
A slice of the  phase space for the non-monotonic 
networks with threshold ${\eta}$ at $T=0$ and $a=1.8$. \\
{\bf Figure 6}\\
Optimal storage capacity ${\alpha}_{c}(a)$ of the non-monotonic network with 
thresholds  ${\eta}=1.0, 0.8$ and $ 0.6$ at $T=0$. \\
{\bf Figure 7}\\
The order parameter $m({\alpha})$ 
at $T=0$ and ${\eta}=0.80$ and $a=3.00$. \\ 
{\bf Figure 8}\\
The order parameter $m({\alpha})$ 
at $T=0$ and ${\eta}=0.80$ and $a=1.60$.


\newpage
\begin{thebibliography}{1}

\bibitem{Hop1}
Hopfield J J {\it Proc. Natl. Acad. Sci. USA} {\bf 79} 2554 (1982) 

\bibitem{Hop2}
Hopfield J J {\it Proc. Natl. Acad. Sci. USA} {\bf 81} 3088 (1984)

\bibitem{Amit}
Amit D J and Gutfreunt H and Sompolinsky H {\it Phys. Rev. Lett.}
 {\bf 55} 1530 (1985)
 
\bibitem{Amit2}
Amit D J and Gutfreunt H and Sompolinsky H {\it Ann. Phys.}
{\bf 173} 30 (1987)


\bibitem{Fon}
Fontanari J F and K$\ddot{\rm o}$berle R {\it J. Physique} {\bf 49} 
13 (1988)

\bibitem{Amari}
Amari S and Maginu K {\it Neural Networks} {\bf 1} 63 (1988)


\bibitem{Nishi2}
Nishimori H and Ozeki T {\it J. Phys. A: Math. Gen.} {\bf 26} 859 (1993)

\bibitem{Gar}
Gardner E {\it J. Phys. A: Math. Gen.} {\bf 20} 3453 (1987)

\bibitem{Gar2}
Gardner E {\it J. Phys. A: Math. Gen.}  {\bf 21} 257 (1988)

\bibitem{Gar3}
Gardner E and Derrida B  {\it J. Phys. A: Math. Gen.}  {\bf 21} 271 (1988)


\bibitem{Ze}
Zertuche F, L$\acute{\rm o}$pez R and Waelbroeck H  {\it J. Phys. A: Math. Gen.} 
{\bf 27} 1575 (1994)

\bibitem{Nishi1}
Nishimori H and Opris I {\it Neural Networks} {\bf 6}
 1061 (1993)

\bibitem{Mori}
Morita M, Yoshizawa S and Nakano K {\it Trans. IEICE} {\bf J73-D-2(2)} 
242 (1990)

\bibitem{Ge}
Geszti T {\it Physical Models of Neural Networks} (World Scientific: 
Singapore)
(1990)

\bibitem{Her}
Hertz J, Krogh A and Palmer R G {\it Introduction to the Theory of Neural 
Computation} (World Scientific: Singapore) Chapter 2

\bibitem{Ed}
Edwards S F and Anderson P W {\it J. Phys. F: Met. Phys.} {\bf 5} 965 (1975)

\bibitem{Mat}
Matus I J and Penez P {\it Phys. Rev. A} {\bf 41} 7013 (1990)

\end{thebibliography}
\end{document}